\pdfoutput=1

\documentclass{jfm}

\usepackage{graphicx}
\usepackage{natbib}
\usepackage{amsmath,amscd,amssymb,dsfont}
\usepackage{bm}
\usepackage{color}
\usepackage{subfigure}

\graphicspath{{figs/}}

\newcommand{\md}{d\kern-0.035cm\char39\kern-0.03cm}
\newcommand{\mD}{D\kern-0.035cm\char39\kern-0.03cm}

\renewcommand{\vec}[1]{\bm{#1}}

\def\be{\begin{equation}}
\def\ee{\end{equation}}
\def\bea{\begin{eqnarray}}
\def\eea{\end{eqnarray}}

\title[Optimal stirring strategies for passive scalar mixing]{Optimal stirring strategies\\ for passive scalar mixing}
\date{}
\author[Z. Lin, J.-L. Thiffeault, and C. R. Doering]%
{Z\ls H\ls I\ns\ls L\ls I\ls N$^1$,\ns
J\ls E\ls A\ls N-L\ls U\ls C\ns\ls T\ls H\ls I\ls F\ls F\ls E\ls A\ls U\ls L\ls T$^{1,2}$\break
\and C\ls H\ls A\ls R\ls L\ls E\ls S\ns\ls R.\ns\ls D\ls O\ls E\ls R\ls I\ls N\ls G$^{1,3}$
}
\affiliation{$^1$Institute for Mathematics and Its Applications, University of
Minnesota, Minneapolis, MN 55455 \\[\affilskip]
$^2$Department of Mathematics, University of Wisconsin, Madison, WI 53706 \\[\affilskip]
$^3$Department of Mathematics, Department of Physics, and Center for
the Study of Complex Systems, University of Michigan, Ann Arbor, MI 48109 \\[\affilskip]
}

\begin{document}
\maketitle

\begin{abstract}
We address the challenge of optimal incompressible stirring to mix an initially inhomogeneous distribution of passive tracers.
As a quantitative measure of mixing we adopt the $H^{-1}$ norm of the scalar fluctuation field, equivalent to the (square-root of the) variance of a low-pass filtered image of the tracer concentration field.
First we establish that this is a useful gauge even in the absence of molecular diffusion: its vanishing as $t \rightarrow \infty$ is evidence of the stirring flow's mixing properties in the sense of ergodic theory.
Then we derive absolute limits on the total amount of mixing, as a function of time, on a periodic spatial domain with a prescribed instantaneous stirring energy or stirring power budget.
We subsequently determine the flow field that instantaneously maximizes the decay of this mixing measure---when such a flow exists.
When no such `steepest descent' flow exists (a possible but non-generic situation) we determine the flow that maximizes the growth rate of the $H^{-1}$ norm's decay rate.
This local-in-time optimal stirring strategy is implemented numerically on a benchmark problem and compared to an optimal control approach using a restricted set of flows. Some significant challenges for analysis are outlined.
\end{abstract}

\section{Introduction.}

The enhancement of mixing by stirring in incompressible flows is an
important phenomenon in the natural sciences and in engineering applications.
A natural question to pose is, how efficient a mixer can an incompressible flow be?
This fundamental question, more precisely posed, is the subject of this paper.

In principle, given an appropriate quantitative measure of mixing along with suitable constraints on the accessible class of flow fields, the most efficient mixing strategy may be determined by solving an optimal control problem.
In practice this may be difficult, so it is useful to consider other approaches that might more easily be implemented, at least theoretically or computationally.
Moreover, it is always useful to know absolute limits on how fast mixing could ever be achieved subject to the relevant constraints.
Such bounds provide a scale upon which particular strategies may be evaluated to gauge their effectiveness.
Here we propose and analyze a theoretical scenario with a particular mixing measure subject to particular constraints where these issues can be investigated analytically and via direct numerical simulation.
As will be seen, there is an interesting gap between the best available {\it a priori} analysis and the simulation results.

The rest of this paper is organized as follows.  In the next section we introduce the basic notions and define the specific problems to be studied.
Rigorous bounds on the rate at which mixing measures can decay for constrained stirring flows are derived in section 3, and an optimal mixing strategy is described in section 4.
In the final section 5 we report computational implementations of the optimal strategy and discuss open challenges suggested by the results.

\section{Problem description.}

Consider the advection of a passive scalar field $\theta(\vec{x},t)$ by a smooth incompressible flow field $\vec{u}(\vec{x},t)$ described by the partial differential equation
\be
\partial_t\theta + \vec{u}\cdot\nabla\theta=0
\label{mastereq}
\ee
along with initial condition $\theta(\vec{x},0)=\theta_0(\vec{x})$.
The stirring field  $\vec{u}$ and the initial distribution $\theta_0$
are periodic in the $d$-dimensional domain $[0,L]^d$ so the subsequent
solution $\theta$ is as well.
Without loss of generality $\theta_0$ and $\theta$ are spatially mean-zero:
\be
\langle\theta(\cdot,t)\rangle \equiv
\frac{1}{L^d}\int_{[0,L]^d}\theta(\vec{x},t)\,d\vec{x}=0.
\ee
We also restrict attention to spatially mean-zero flows, i.e., $\langle u_i(\cdot,t) \rangle = 0$ for $i=1,\dots,d$.

The goal of effective stirring is to redistribute the passive tracer density to achieve a maximal rate of mixing as quantified by the relevant mixing measure.
To gauge the effectiveness of the stirring we adopt as a mixing measure the $H^{-1}$ norm on mean-zero functions where,
for any real parameter $a$, the $H^{-a}$ norm is a weighted sum of Fourier coefficients of the scalar field:
\be
\|\theta(\cdot,t)\|_{H^{-a}}^2 \equiv \| | \nabla |^{-a}\theta(\cdot,t)\|_{L^2}^2 =
 \sum_{\vec{k}\neq
   0}{k^{-2a}}{|\hat{\theta}_{\vec{k}}(t)|^2}
\label{def_h_1}
\ee
where $k = |\vec{k}|$ and
\be
\hat{\theta}_{\vec{k}}(t) =  \frac{1}{L^{d/2}} \int_{[0,L]^d} e^{-i \vec{k} \cdot \vec{x}} \theta(\vec{x},t) \, d\vec{x}.
\ee
The operator $|\nabla|^{-a}$ generally acts in Fourier space as multiplication by $k^{-a}$ although when $a$ is an odd integer $\nabla^{-a}$ is naturally defined as multiplication by $-{i\vec{k}}/{k^{(a+1)}}$.

We focus in particular on the $H^{-1}$ norm which is related to the large-scale mixing measure previously studied by a subset of the authors~\citep{DoeringThiffeault2006,Shaw2007}.
It measures the variance of a low-pass-filtered image of the concentration field: the smaller the $H^{-1}$ norm is, the more homogeneous the scalar field is on large spatial scales.
In many applications molecular diffusion, implemented mathematically by an additional $\kappa \Delta \theta$ term on the right hand side of the equation (\ref{mastereq}), dissipates the variance, i.e., the $H^{0}$ norm, of the scalar field.
But even in the absence of molecular diffusion when the $H^{0}$ norm is conserved, or when the diffusion is ineffective on the length and time scales of interest, the relatively large-scale structures in the scalar field may nevertheless decay when a `mix norm' like the $H^{-1}$ norm is employed.
This idea was introduced by \cite{Mathew2005} for the $H^{-1/2}$ norm and is extended to other norms here.

For $a > 0$ the norms $H^{-a}$ provide a quantitative measure of mixing in the sense of ergodic theory.
To see this consider Lagrangian particle trajectories $\vec{X}(t)$ defined by
\be
\frac{d\vec{X}}{dt} = \vec{u}(\vec{X}(t),t)
\ee
with random initial condition $\vec{X}(0)$ distributed according to the density $\rho_0(\vec{x})$.
Then the tracer particle positions are distributed according to the solution $\rho(\vec{x},t)$ of
\be
\partial_t \rho + \vec{u} \cdot \nabla \rho=0 \ee with initial condition $\rho_0(\vec{x})$.
Incompressible advection conserves the variance of $\rho(\vec{x},t)$: the mean-zero field $\theta = \rho - L^{-d}$ also satisfies the advection equation, and multiplying (\ref{mastereq}) by $\theta$, integrating over the domain, and integrating by parts yields~${\text{d}\|\theta\|_{L^2}}/{\text{d}t} = 0$.
Hence the variance of the density does not measure mixing.
Rather, the stirring field $\vec{u}(\vec{x},t)$ is called {\it mixing} if for every periodic square-integrable function $g(\vec{x})$ on $[0,L]^d$,
\be \lim_{t \rightarrow \infty} \ \int_{[0,L]^d} g(\vec{x}) \rho(\vec{x},t)\,d\vec{x} \ = \ \langle g \rangle.
\ee
See, for example, \cite{Lasota}.
The utility of the $H^{-a}$ norms (\ref{def_h_1}) are indicated by the
following theorem, an extension of \cite{Mathew2005}.

\smallskip
\noindent
{\bf Theorem:}  Suppose the spatially mean-zero function $\theta(\vec{x},t)$ is bounded uniformly in $L^2([0,L]^d)$ for all $t>0$.
Then
\[
\lim_{t \rightarrow \infty} \ \int_{[0,L]^d} g(\vec{x})\,
\theta(\vec{x},t)\, d\vec{x} = 0 \ \ \forall g \in L^2
\ \ \Longleftrightarrow \ \  \lim_{t \rightarrow \infty} \ \|\theta(\cdot,t)\|_{H^{-a}} = 0,\text{\ for any\ } a>0.
\]

\noindent
See the appendix for an elementary proof that applies as well to many other measures which could serve just as effectively as a mix norm in this regard.

The upshot is that more rapid self-averaging characteristic of the intuitive notion of mixing is indicated by more rapid decay of the $H^{-a}$ norm.
Alternative measures have also been used to characterize mixing and the associated control problem.  See, for example, ~\cite{Sharma1997, DAlessandro1999, Vikhansky2002, Schumacher2003b, Balogh2005, Mathew2005, Thiffeault2006, Mathew2007, Constantin2008, Cortelezzi2008, Thiffeault2008, Gubanov2010}.
The $H^{-1}$ norm adopted here, however, allows for the development of a particularly straightforward and operational stirring strategy.

Constraints must be imposed upon on the available flow fields in order to formulate an optimization problem.
We focus on constraints of bounded instantaneous kinetic energy, proportional to the $L^2$ norm of the velocity $\|\vec{u}(\cdot,t)\|_{L^2}^2$, or bounded instantaneous power dissipation in the flow, which for Newtonian fluids with periodic boundary conditions is proportional to the $H^1$ norm of the velocity $\|\nabla\vec{u}(\cdot,t)\|_{L^2}^2=\sum_{i,j=1}^d \| \partial_i u_j(\cdot,t)\|_{L^2}^2$.
That is, we consider flow fields $\vec{u}$ satisfying either
\be
\int_{[0,L]^d}|\vec{u}(\vec{x},t)|^2 \text{d} \vec{x} \ = U^2 \, L^d
\label{fixecons}
\ee
or
\be
\int_{[0,L]^d}|\nabla\vec{u}(\vec{x},t)|^2\text{d} \vec{x}  \ = \int_{[0,L]^d} \frac{1}{4} \sum_{i,j=1}^d (\partial_i u_j+\partial_j u_i)^2\text{d} \vec{x}  \ = \ \frac{L^d}{\tau^2}
\label{fixpcons}
\ee
defining, respectively, the root-mean-square velocity
$U=\langle|\vec{u}|^2\rangle^{1/2}$ or rate of strain $\tau^{-1}=\langle|\nabla \vec{u}|^2\rangle^{1/2}$ of the stirring.
Given either constraint two natural questions are:
\renewcommand{\theenumi}{\Roman{enumi}}

\smallskip
\begin{enumerate}
\item What flow \emph{minimizes} the mixing measure evaluated at a final time $t_{\text{fin}}>0$?
\item What flow \emph{maximizes} the instantaneous decay rate of the mixing measure?
\end{enumerate}
\smallskip
These questions have different answers implying different `optimal' stirring strategies.
\cite{Mathew2007} studied problem (I) and solved it numerically for a limited set of flow field configurations using the $H^{-1/2}$ norm as the mixing measure.
They computed the controls in the form of a time varying linear combination of two simple cellular flows.
That approach is global in time since it requires keeping track of the complete evolutionary history of the system within the interval $[0,t_{\text{fin}}]$.

In this paper we address question (II) and consider flows that produce the steepest descent of the $H^{-1}$ mixing measure at each instant in time.
This local-in-time strategy identifies an optimal mixing flow $\vec{u}(\vec{x},t)$ at time $t$ using only a snapshot of the scalar field $\theta(\vec{x},t)$ at that instant.
It should be noted that for the optimization problem (I) seeking to minimize the mixing measure at a final time, natural constraints might also be the total action (the integral of $\|\vec{u}(\cdot, t)\|_2^2$ from $t=0$ to $t_{\text{fin}}$) or the total energy (proportional to the time integral of $\|\nabla\vec{u}(\cdot, t)\|_2^2$ from $t=0$ to $t_{\text{fin}}$).

\section{Absolute bounds on mixing rates.}

As a starting point it is useful to identify absolute limits on the rate at which scalar fields might be mixed by fluids satisfying the instantaneous energy or power constraints.
Toward this end we note that incompressible advection conserves not only the variance of $\theta$ but also that the (weak) maximum principle ensures that
the $L^{\infty}$ norm (the supremum of $|\theta|$ over the spatial domain) is conserved as well, i.e., $\|\theta(\cdot, t)\|_{L^{\infty}} =  \|\theta_0\|_{L^{\infty}}$ at every time $t>0$.

First consider the fixed energy constraint (\ref{fixecons}).
Multiplying (\ref{mastereq}) by $-\Delta^{-1}\theta$, integrating over the spatial domain, and integrating by parts implies
\be
\frac{\text{d}}{\text{d}t} \|\nabla^{-1}\theta\|_{L^2}^2 = \frac{\text{d}}{\text{d}t} \|\theta\|_{H^{-1}}^2 = -2\int \theta \, \vec{u}\cdot\nabla (\Delta^{-1} \theta)\, d\vec{x}.
\label{bnd1}
\ee
The H\"older and Cauchy--Schwarz inequalities then give
\be
\frac{\text{d}}{\text{d}t} \|\theta\|_{H^{-1}}^2 \ge -2 \,\|\vec{u}\|_{L^2} \, \|\theta\|_{L^{\infty}}  \, \|\theta\|_{H^{-1}}=-2\ U  L^{d/2}\, \|\theta_0\|_{L^{\infty}} \,\|\theta\|_{H^{-1}},
\label{bnd1a}
\ee
and dividing both sides by $2\,\|\theta\|_{H^{-1}}$ and integrating in time yields
\begin{eqnarray}
\|\theta(\cdot,t)\|_{H^{-1}} \ \ge \ \|\theta_0\|_{H{-1}} \ - \ \ U \, L^{d/2} \, \|\theta_0\|_{L^{\infty}} \ t.
\label{bnd3}
\end{eqnarray}
This rigorous estimate does not rule out perfect mixing as measured by the $H^{-1}$ norm after a finite time, but it does bound the absolute minimum mixing time from below by
\be
t_{\text{mix}} \ = \ \frac{1}{U L^{d/2}} \, \frac{\|\theta_0\|_{H^{-1}}}{\|\theta_0\|_{L^{\infty}}} \ = \ \frac{l_0}{2 \pi U},
\ee
singling out a length scale
\be
l_0 = 2 \pi \frac{\langle |\nabla^{-1}\theta_0|^2 \rangle^{1/2}}{\|\theta_0\|_{L^{\infty}}} \ (\le L \ \text{due to the Poincar\'e and H\"older inequalities)}
\ee
characterizing the spatial extent or `size' of initial inhomogeneities in the scalar field.
The lower bound in (\ref{bnd3}) simply states that under the constant energy constraint, the time it takes to achieve complete mixing is at least as long as the time it takes to transport scalar inhomogeneities across the characteristic distance $\sim l_0$ at the rms speed $U$ of the flow.
Whether or not this limiting mixing rate can actually be achieved, or even approached, by any suitably constrained stirring flow remains to be seen.

The analysis is rather different for flows subject to the fixed power constraint (\ref{fixpcons}).
For convenience we define the filtered scalar field
\be
\varphi(\vec{x},t) = \left( \Delta^{-1} \theta \right)(\vec{x},t) =
- \sum_{\vec{k}\neq 0} e^{i \vec{k}\cdot\vec{\theta}} {k^{-2}} \hat{\theta}_{\vec{k}}(t).
\ee
Then starting from (\ref{bnd1}) more integrations by parts yield
\be
\frac{\text{d}}{\text{d}t} \|\theta\|_{H^{-1}}^2 \ = \ \frac{\text{d}}{\text{d}t} \|\nabla \varphi\|_{L^2}^2
\ = \ -2 \sum_{i, j = 1}^d \int \varphi \, \frac{\partial u_i}{\partial x_j} \,  \frac{\partial^2 \varphi}{\partial x_i \partial x_j} \text{d} \vec{x}
\label{bndSTEP}
\ee
and the H\"older and Cauchy--Schwarz inequalities imply
\be
\frac{\text{d}}{\text{d}t} \|\theta\|_{H^{-1}}^2 \ \ge \ -2 \, \|\varphi\|_{L^{\infty}} \, \|\nabla\vec{u}\|_{L^2} \, \|\Delta \varphi\|_{L^2}
\ = \ -2 \, \|\varphi\|_{L^{\infty}} \, \frac{L^{d/2}}{\tau} \, \|\theta_0\|_{L^2}.
\label{bnd4}
\ee
In order to close the differential inequality it is necessary to bound the sup norm $\|\varphi\|_{L^{\infty}}$ in terms of the $H^{+1}$ norm of $\varphi$, i.e., $\|\nabla\varphi\|_{L^2} = \|\nabla^{-1}\theta\|_{L^2}=\|\theta\|_{H^{-1}}$, and some conserved (or otherwise {\it a priori} bounded) quantities.
This is possible in $2$ and $3$ spatial dimensions.

First consider $d=3$.
For mean-zero functions on the 3-torus, there exists an  ${\cal O}(1)$ pure number $C_3$ so that
\be
\|\varphi\|_{L^{\infty}} \ \le \ C_3 \, \|\nabla \varphi\|_{L^2}^{1/2} \, \|\Delta \varphi\|_{L^2}^{1/2} \ = \ C_3 \, \|\theta\|_{H^{-1}}^{1/2} \, \|\theta_0\|_{L^2}^{1/2}.
\label{Ag3}
\ee
For an elementary proof of the inequality see \cite{DG}.
Thus
\be
\frac{\text{d}}{\text{d}t} \|\theta\|_{H^{-1}}
\ \ge \ - \, \frac{C_3 \, L^{3/2}}{\tau} \, \|\theta_0\|_{L^2}^{3/2} \,  \|\theta\|_{H^{-1}}^{-1/2}
\ee
and
\be
\|\theta(\cdot,t)\|_{H^{-1}} \ \ge \ \|\theta_0\|_{H^{-1}} \, \left[ \, 1 \ - \frac{3 \, C_3}{2 \, \tau} \,
 \left( \frac{L \, \|\theta_0\|_{L^{\infty}}}{\|\theta_0\|_{H^{-1}}}\right)^{3/2}  \times t \, \right]^{\frac{2}{3}}
\ee
as long as the term in brackets is non-negative.
Again the rigorous analysis does not rule out perfect mixing in a finite time but bounds the minimal mixing time from below by
\be
t_{\text{mix}} \ = \ \tau \times \frac{2}{3 \, C_3} \, \left(\frac{\ell_0}{2 \pi L}\right)^{\frac{3}{2}} ,
\label{TMP3}
\ee
where
\be
\ell_0 \ =2 \pi  \frac{\|\theta_0\|_{H^{-1}}}{\|\theta_0\|_{L^{2}}} \ (\le \ L \ \text{by Poincar\'e's inequality)}
\label{LP}
\ee
is another length scale characterizing the size of inhomogeneities in the initial distribution.

The linear dependence of $t_{\text{mix}}$ on $\tau$ is not surprising.
This rigorous estimate is, however, more than na\"ive dimensional analysis because the lower bound on the mixing time allows for a nontrivial $L$-dependence.
The minimal mixing time estimate in (\ref{TMP3}) suggests that the availability of larger {\it domain} length scales $L$ for the flow to access may in fact facilitate mixing, perhaps by allowing for more effective `folding' to accompany judiciously localized `stretching'.
The suggestion (which remains a conjecture at this point) is that if
$N^3$ copies of the same initial distribution with basic scale
$\ell_0$ are assembled into a larger periodic domain $[0,N \times
L]^3$, the scalar might possibly be mixed faster by a flow with the
same rms rate of strain than on the elementary cell $[0,L]^3$.

A similar sort of system size dependence is suggested in two dimensions.
The $d=2$ analog of (\ref{Ag3}) is (again, see \cite{DG} for an elementary derivation)
\[
\|\varphi\|_{L^{\infty}} \ \le \ C_2 \, \|\nabla \varphi\|_{L^2}
\sqrt{1 + \log{\left[ \frac{L \, \|\Delta \varphi\|_{L^2}}{2 \pi \|\nabla \varphi\|_{L^2}} \right]}} \ = \
C_2 \, \|\theta\|_{H^{-1}}
\sqrt{1 + \log{\left[ \frac{L \, \|\theta_0\|_{L^2}}{2 \pi  \|\theta\|_{H^{-1}}} \right]}}
\]
for an ${\cal O}(1)$ constant $C_2$.
Thus
\be
\frac{\text{d}}{\text{d}t} \|\theta\|_{H^{-1}}
\ \ge \ - \, \frac{C_2 \, L}{\tau} \, \|\theta_0\|_{L^2} \, \sqrt{1 + \log{\left[ \frac{L \, \|\theta_0\|_{L^2}}{2 \pi  \|\theta\|_{H^{-1}}} \right]}}.
\ee
This differential inequality also does not prevent $ \|\theta\|_{H^{-1}}$ from vanishing in finite time, but it guarantees that $ \|\theta\|_{H^{-1}}$ cannot vanish before the absolute minimum mixing time
\be
t_{\text{mix}} \ = \ \tau \, \frac{e}{2 \pi C_2} \, \int_{\log{\frac{L}{\ell_0}}}^{\infty} \frac{e^{-\zeta}}{\sqrt{\zeta}} \, \text{d}\zeta
\ee
where $\ell_0$ is defined in (\ref{LP}).
For $L \gg \ell_0$ the minimum mixing time is
\be
t_{\text{mix}} \sim \tau \times \frac{\ell_0}{L} \times \left(\log{\frac{L}{\ell_0}}\right)^{-1/2},
\label{TMP2}
\ee
again allowing for the possibility that it may take less time to mix
within a larger volume.

In contrast to these potentially finite-time mixing scenarios, if the flow field is constrained to have a {\it uniformly} bounded (in space and time) rate of strain then it can decay at most exponentially.
In any spatial dimension if $\|(\nabla \vec{u})_{\text{sym}}\|_{L^{\infty}} \le \gamma < \infty$ then
\be
\frac{\text{d}}{\text{d}t} \|\theta\|_{H^{-1}}^2 = 2 \int \nabla^{-1} \theta \cdot (\nabla \vec{u}) \cdot \nabla^{-1} \theta\, d\vec{x}
\ \ge \ -2 \gamma \, \|\theta\|_{H^{-1}}^2
\ee
and Gr\"{o}nwall's inequality implies an exponential lower bound on the mixing measure:
\be
\|\theta(\cdot,t)\|_{H^{-1}} \ \ge \ \|\theta_0\|_{H^{-1}} \, e^{-\gamma \, t}.
\ee
Thus if the rate of strain is bounded then the mixing rate cannot increased by increasing the domain scale (holding all other constraints fixed).
Therefore if there is any real precision regarding the $L$-dependence
in the estimates (\ref{TMP3}) or (\ref{TMP2}) then it depends
crucially on a fixed-power flow's freedom to locally intensify the strain.

\section{Optimal stirring.}

Recalling (\ref{bnd1}) we write
\be
\frac{\text{d}}{\text{d}t}\|\nabla^{-1}\theta\|^2_2=-2\int \vec{u}\cdot(\theta\nabla\varphi)\text{d}\vec{x}
=-2\int \vec{u}\cdot\mathds{P}(\theta\nabla\varphi)\text{d}\vec{x}
\label{mainfcnal}
\ee
where $\mathds{P}(\cdot)$ is the projector onto divergence-free fields defined by
\be
\mathds{P}(\vec{v})  = \vec{v} - \nabla \Delta^{-1} (\nabla \cdot \vec{v}).
\ee
Then with either the fixed energy (\ref{fixecons}) or fixed enstrophy
(\ref{fixpcons})  constraints it is easy to see that the optimal mixer
maximizing the rate of decay of the mix-norm $H^{-1}$ is
\be
\vec{u}_e=U \, \frac{\mathds{P}(\theta\nabla\varphi)}{\langle |\mathds{P}(\theta\nabla\varphi)|^2 \rangle^{1/2}}
\label{elfixesol}
\ee
or
\be
\vec{u}_p=\frac{1}{\tau}\,\frac{-\Delta^{-1}\mathds{P}(\theta\nabla\varphi)}{\langle |\nabla^{-1}\mathds{P}(\theta\nabla\varphi)|^2 \rangle^{1/2}}
\label{elfixpsol}
\ee
{\it if} the norm in the denominator does not vanish.
So long as the relevant optimal stirring exists, the ideal instantaneous stirring strategy is to implement it at each instant of time.
But if either of the norms in the denominators vanishes then
$\mathds{P}(\theta\nabla\varphi) = 0$ throughout the domain and {\it
  no} incompressible flow can instantaneously decrease the mix-norm $H^{-1}$.

A sufficient (but not to our knowledge necessary) condition for such degeneracies is that the scalar field $\theta$ satisfies
\be
\Delta \theta=F(\theta),
\ee
which includes cases where $\theta$ is an eigenfunction of the Laplacian.
Such situations arise naturally as convenient initial conditions but we suspect that they are non-generic distributions among solutions of the advection equation.
Nevertheless when and if this situation develops some other strategy must be adopted to stir the fluid.

The natural thing to do when $\mathds{P}(\theta\nabla\varphi) = 0$ is find the flow that maximizes the rate of increase of the rate of decrease of the norm.
That is, seek the flow that minimizes
\be
\frac{\text{d}^2}{\text{d} t^2}\|\theta\|^2_{H^{-1}} = 2\int\left[\vec{u}\cdot\nabla\varphi \, \nabla\theta \cdot\vec{u} - (\vec{u}\cdot\nabla\theta)\Delta^{-1}(\vec{u}\cdot\nabla\theta)\right]\, \text{d}\vec{x}.
\ee
The optimal incompressible $\vec{u}$ here solves the eigenvalue problem
\be
\lambda\vec{u}=\mathds{P}\Big((\vec{u}\cdot\nabla\theta)\nabla\varphi+(\vec{u}\cdot\nabla\varphi)\nabla\theta-2[\Delta^{-1}(\vec{u}\cdot\nabla\theta)]\nabla\theta\Big)
\label{eigfixe}
\ee
for fixed energy constraint or
\be
\lambda\vec{u}=-\Delta^{-1} \mathds{P}\Big((\vec{u}\cdot\nabla\theta)\nabla\varphi+(\vec{u}\cdot\nabla\varphi)\nabla\theta-2[\Delta^{-1}(\vec{u}\cdot\nabla\theta)]\nabla\theta\Big)
\label{eigfixp}
\ee
for fixed power constraint.
In either case we seek the eigenvector (field) corresponding to the
minimum eigenvalue $\lambda_- < 0 $ to use as the stirring field momentarily until $\mathds{P}(\theta\nabla\varphi) \ne 0$.

These are generally difficult eigenvalue problems.
To make some analytical progress consider the special case $\theta_0(\vec{x})=\sin(kx)$ that we will use as an initial condition in the computational test reported in the next section in spatial dimension $d=2$.
Introducing the stream function so that $u=\partial \psi / \partial y$ and $v=-\partial \psi / \partial x$ and taking the curl of (\ref{eigfixe}),
\be
-\lambda\Delta\psi=-2\nabla\times\mathds{P}\vec{v}=-2\nabla\times(\vec{v}-\nabla\Delta^{-1}(\nabla\cdot\vec{v}))=-2\nabla\times\vec{v}=2\frac{\partial v_x}{\partial y}
\label{eigfixepsi}
\ee where
$\vec{v}(\vec{x},t)=(\vec{u}\cdot\nabla\theta)\nabla\varphi+(\vec{u}\cdot\nabla\varphi)\nabla\theta-2[\Delta^{-1}(\vec{u}\cdot\nabla\theta)]\nabla\theta$.
Writing $\psi(x,y)=\sum_{m,n}\hat{\psi}_{mn}e^{2\pi i(mx+ny)/L}$ produces a matrix equation for the Fourier coefficients $\hat{\psi}_{mn}$.
With the fixed energy constraint (\ref{fixecons}) finite-resolution ($n, \, m \le N$) numerical solutions to this matrix equation show that the
minimum eigenvalue is always associated with the $N^{\text{th}}$-harmonic, suggesting that the optimal flow depends on the imposed Fourier cutoff.
Focusing instead on the fixed enstrophy constraint (\ref{fixpcons}), it is readily shown that to leading order in the initial data wavenumber $k$ the minimum eigenvalue we seek is
\be
\min_{m,n}\lambda_{mn}(k)=\lambda_{m^*,1}(k)\sim
-\frac{1}{k^2}\left[1+{\cal O}\left( (kL)^{-4}\right)\right]
\label{gen2varsolspeceigv}
\ee
and the associated eigenfunction, modulo an arbitrary phase shift in $y$, is approximately proportional to the first harmonic, i.e.,
$u \propto \cos(2\pi y/L)$ and $v = 0.$

Finally, we remark that unless the scalar field is a superposition of the lowest available modes it can be {\it unmixed}.
Indeed, when $\mathds{P}(\theta\nabla\varphi) \ne 0$ then simply reversing the optimal flow momentarily unmixes the scalar as defined by the $H^{-1}$ norm, and the same is true in the exceptional situations with $ \mathds{P}(\theta\nabla\varphi) \equiv 0$.

\section{Computational tests and discussion.}

\begin{figure}
  \centering
  \subfigure[]{
    \includegraphics[height=5cm]{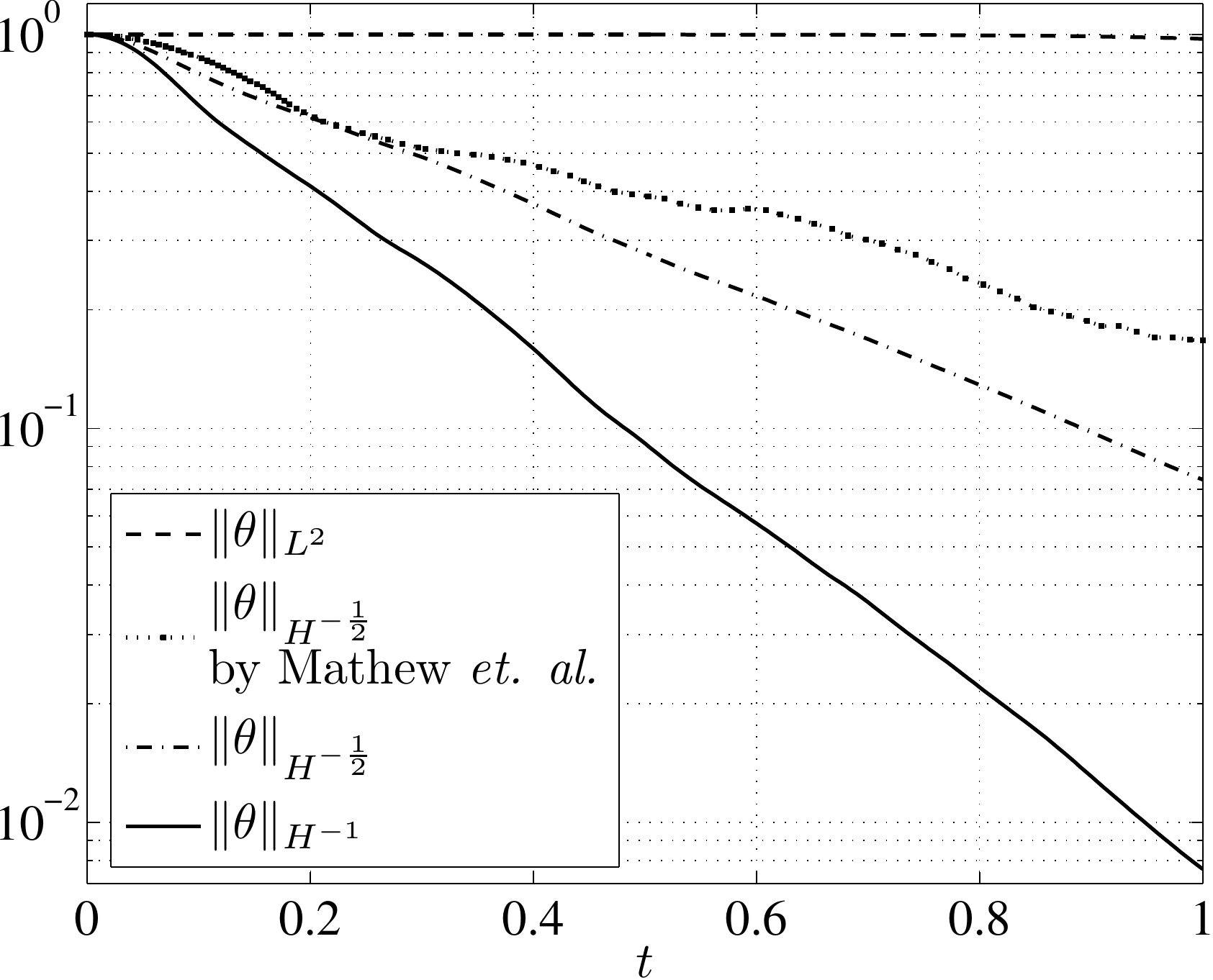}
    \label{OptMix2}
  }
  \hskip 0.5cm
  \subfigure[]{
 \includegraphics[height=5cm]{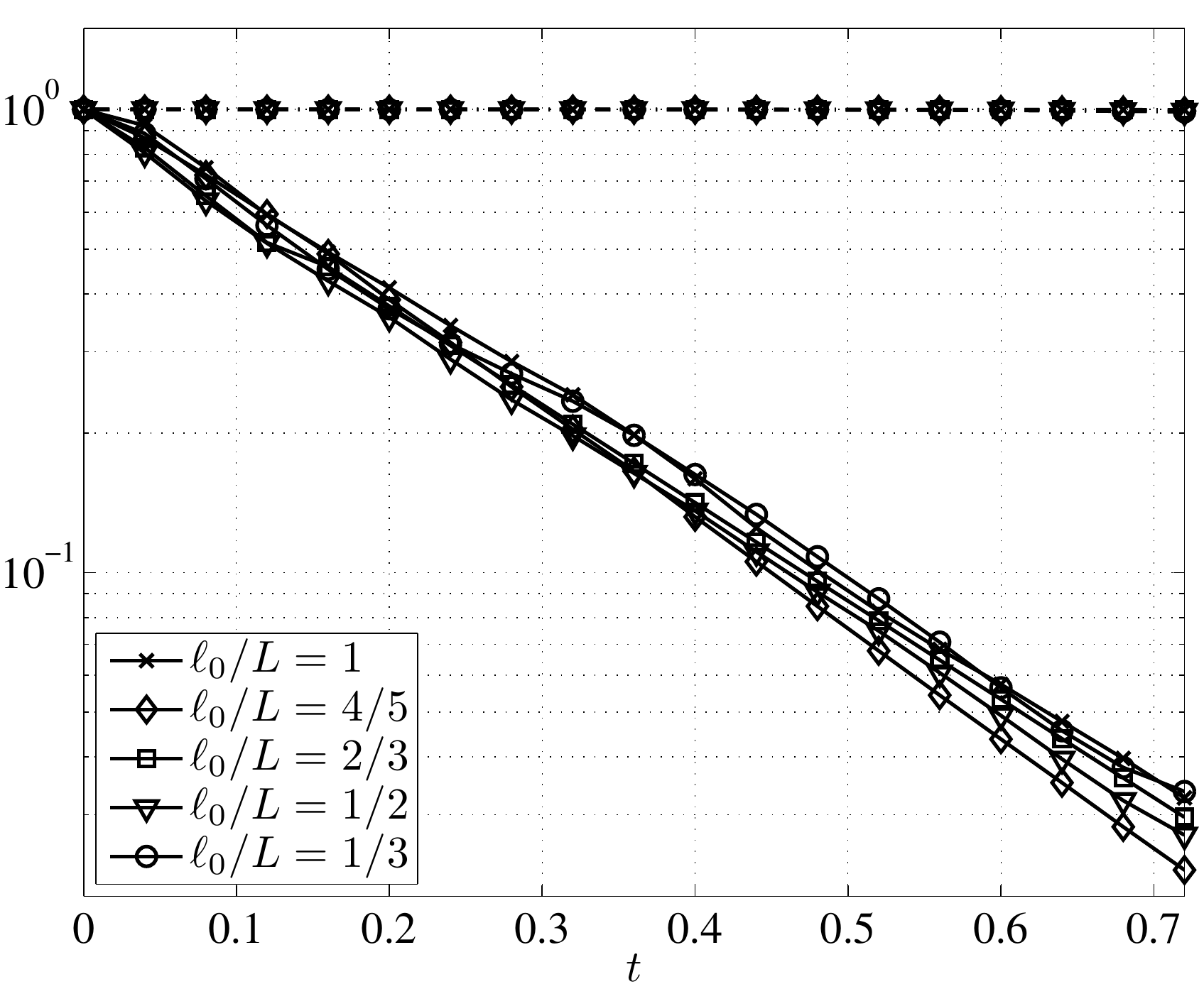}
    \label{H_1atDiffl0}
  }
  \caption{(a) Evolution of mixing measures for
    $\theta_0(\vec{x})=\sin(x)$.  All norms are rescaled by their
    initial values.  (b) Decay of $H^{-1}$ for different values of
    $\ell_0/L$ (solid lines).  The $\ell_0/L=1$ case uses $\theta_0=\sin x$, while
    the $\ell_0/L=4/5$ initial distribution is a sum of 5
    low-wavenumber harmonics, $\theta_0=\sin x + 1.92 (-0.542 \cos 2x
    + 0.8267 \sin y - 0.6592 \cos 2y + 0.3998 \sin x \cos y + 0.6516
    \cos x \sin y)$.  The $\ell_0/L=2/3$ data used $\theta_0(x)=\sin
    x+\sqrt{5/3} \sin 3y$.  The $\ell_0/L=1/2$ case uses
    discontinuous (non-periodic) initial data, $\theta_0 =
    \sin[2(1+0.5\sin y) x]$, and the $\ell_0/L= 1/3$ initial data is
    $\sin 3x$. The dot-dash lines illustrates the conservation of $L^2$ norms with different data in the simulations.}
\end{figure}

The first test of the optimal stirring strategy uses initial scalar distribution $\theta_0(x)=\sin x$ (with $\ell_0/L=1$) in a domain of size $L=2 \pi$ in $d=2$ spatial dimensions.
We implemented the fixed power constraint with $\tau^{-1}=6.25\times(2\pi)^2$, which is equivalent to the amplitude of the bi-component control utilized by \cite{Mathew2007} .
Figure \ref{OptMix2} shows the evolution of the scalar mix norms.
With the optimal mixing protocol proposed here, after an initial transient where the flow is chosen to maximize $\frac{\text{d}^2}{\text{d} t^2}\|\theta\|^2_{H^{-1}}$, a robust exponential decay of the $H^{-1}$ norms emerges.
In order to compare with the two-component flow optimal control results of \cite{Mathew2007} we also computed the $H^{-1/2}$ norm that they utilized and reported.
Not unexpectedly, expanding the set of available flows from two possibilities to everything within the constant power constraint allows for faster mixing.
Nevertheless the local-in-time optimization strategy employed here leaves room for further improvement.
The conserved $L^2$ norm was also monitored as a numerical check.

\begin{figure}
\centering
\includegraphics[height=8.0cm]{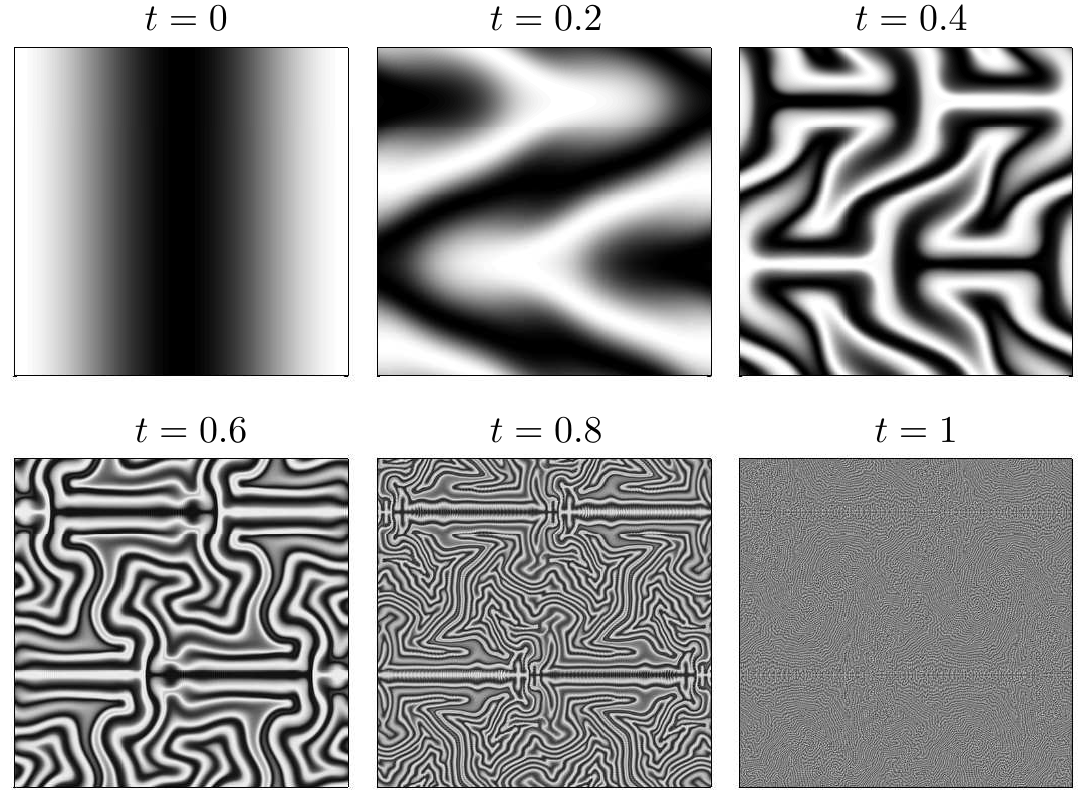}
\caption{\label{OptMixEvol}Evolution of scalar field in $[0,2\pi]^2$ with optimal
  mixing strategy (\ref{elfixpsol}) and solution to (\ref{eigfixp})
  with fixed power and $\theta_0(\vec{x})=\sin
  x$.}
\end{figure}

To check for a possible dependence of the mixing rate on $\ell_0/L$ we also considered different initial data, and the results are displayed in Figure \ref{H_1atDiffl0}.
As is evident, the mixing scheme generates robust exponential decay of the mix-norm.
Exponential decay at a rate independent of the initial data is the sort of dynamics that might be expected for a uniform boundedness constraint on the rate of strain, but not the finite-time mixing that the best available analysis allows for when only the root-mean-square rate of strain is bounded.
It is possible that the rigorous analysis may only be saturated by a global-in-time, i.e., optimal control, stirring strategy but this discrepancy between rigorous analysis and simulation constitutes a major open question.
Figure \ref{OptMixEvol} shows the snapshots of the scalar field
evolution with initial distribution $\theta_0(x)=\sin x$ under the
local fixed-power optimal mixer.  The optimal flow generates a suggestively self-similar cascade of the scalar fluctuations to small scales.  Other initial conditions generate similar cascades.

\section{Acknowledgments}
The authors gratefully acknowledge discussions with Igor Mezi\'c.
This work was supported by NSF Awards PHY-0555324 \& PHY-0855335 (CRD) and DMS-0806821 (J-LT).
This research was initiated at the Optimal Transport program at NSF's Institute for Pure and Applied Mathematics (UCLA) and developed further within the Geophysical Fluid Dynamics Program at Woods Hole Oceanographic Institution, supported by NSF and ONR.


\appendix

\section*{Appendix: Proof of theorem}

\newcommand{\mathnotation}[2]{\newcommand{#1}{\ensuremath{#2}}}

%
%
\newcommand{\norm}[1]{\l\lVert#1\r\rVert}
\newcommand{\Ltnorm}[1]{\norm{#1}_{L^2}}

%
%
\renewcommand{\l}{\left}            
\renewcommand{\r}{\right}           
\renewcommand{\time}{t}             
\mathnotation{\pd}{\partial}        
\mathnotation{\grad}{\nabla}        
\renewcommand{\div}{\nabla\cdot}    
\mathnotation{\lapl}{\Delta}        
\mathnotation{\mlapl}{(-\Delta)}    
\mathnotation{\imi}{i}
\mathnotation{\dint}{\,{\mathrm{d}}}

\mathnotation{\xc}{x}               
\mathnotation{\xv}{\vec{\xc}}        
\mathnotation{\kc}{k}               
\mathnotation{\kv}{{\vec{\kc}}}      
\mathnotation{\km}{\kc}             
\mathnotation{\Lsc}{L}              
\mathnotation{\Vol}{[0,\Lsc]^d}     
\mathnotation{\dV}{\,d\vec{x}}

\mathnotation{\qq}{a}               
\mathnotation{\fw}{\theta}          
\mathnotation{\gt}{g}               
\mathnotation{\T}{T}                
\mathnotation{\Sobo}{H}             
\mathnotation{\Cu}{C}               
\mathnotation{\NK}{K}               
\mathnotation{\eps}{\epsilon}       

We use the norm~\eqref{def_h_1} for mean-zero functions, $\norm{\fw}_{\Sobo^{-\qq}} = \bigl(\sum_\kv \km^{-2\qq} \lvert \hat\fw_\kv\rvert^2\bigr)^{1/2}$, although the proof easily generalizes to other norms.
Suppose that~$\fw(\cdot,\time)$ is uniformly bounded in~$L^2(\Vol)$, so that~$\Ltnorm{\fw(\cdot,\time)} \le \Cu$, and~$\lim_{\time\rightarrow\infty}\norm{\fw(\cdot,\time)}_{\Sobo^{-\qq}} \rightarrow 0$ for some~$\qq>0$.  Then for any~$\gt\in L^2(\Vol)$,
\begin{align*}
  \biggl\lvert\int_{\Vol} \fw\,\gt\dV\biggr\rvert
  &=
  \biggl\lvert
  \sum_{\km \le \NK}
  {\km^{-\qq}}\,\hat\fw_\kv\,
  {{\km^{\qq}}}{\hat\gt_\kv^*}
  + \sum_{\km > \NK} \hat\fw_\kv\,\hat\gt_\kv^*
  \biggr\rvert \\
  &\le
  \norm{\fw}_{\Sobo^{-\qq}}\,
  \biggl(\sum_{\km \le \NK}
  {\km^{2\qq}}{\lvert\hat\gt_\kv\rvert^2}\biggr)^{1/2}
  + \Ltnorm{\fw}
  \biggl(\sum_{\km > \NK} \lvert\hat\gt_\kv\rvert^2\biggr)^{1/2}.
\end{align*}
Given~$\eps>0$, first choose~$\NK(\eps)$ such that $\bigl(\sum_{\km >
  \NK(\eps)} \lvert\hat\gt_\kv\rvert^2\bigr)^{1/2} \le {\eps}/{2\Cu}$,
then choose~$\T(\eps)$ such that
$\norm{\fw(\cdot,\T(\eps))}_{\Sobo^{-\qq}} \le
\tfrac12\eps\,\bigl(\sum_{\km \le \NK(\eps)}
{\km^{2\qq}}{\lvert\hat\gt_\kv\rvert^2}\bigr)^{-1/2}$, for $\time >
\T(\eps)$.
Then
\begin{equation*}
  \biggl\lvert\int_{\Vol} \fw\,\gt\dV\biggr\rvert
  \le
  \tfrac12\l(1 + \Cu^{-1}\Ltnorm{\fw}\r)\eps \le \eps,
  \qquad \time > \T(\eps),
\end{equation*}
which implies that~$\fw$ converges weakly to zero
as~$\time\rightarrow\infty$.  (This is true even for~$\qq=0$.)

Conversely, suppose~$\Ltnorm{\fw(\cdot,\time)} \le \Cu$ for
all~$\time$ and~$\lim_{\time\rightarrow\infty}\int_{\Vol}\fw
\gt\dV\rightarrow 0$ for all~$\gt\in L^2(\Vol)$.  By
choosing~$\gt=\exp(-\imi\kv\cdot\xv)$ we see that all the Fourier
coefficients~$\hat\fw_\kv(\time)\rightarrow 0$
as~$\time\rightarrow\infty$.
Also, because~$\Ltnorm{\fw(\cdot,\time)}^2 = \sum_\kv\lvert\hat\fw_\kv(\time)\rvert^2\le\Cu^2$, each~$\lvert\hat\fw_\kv(\time)\rvert \le \Cu$ for all~$\time$.

Thus
\begin{equation*}
  \norm{\fw}_{\Sobo^{-\qq}}^2
  =
  \sum_{\km \le \NK} \km^{-2\qq}
  \lvert \hat\fw_\kv\rvert^2
  + \sum_{\km > \NK} \km^{-2\qq}
  \lvert \hat\fw_\kv\rvert^2
  \le
  \sum_{\km \le \NK} \km^{-2\qq}
  \lvert \hat\fw_\kv\rvert^2
  + \NK^{-2\qq} \Ltnorm{\fw}.
  \label{eq:laststep}\tag{A.1}
\end{equation*}
For any~$\eps>0$ choose~$\NK(\eps) \ge (2C/\eps)^{1/2a}$ so that~$\km^{-2\qq} \Ltnorm{\fw} \le \km^{-2\qq}\Cu < \eps/2$ for~$\km\ge\NK(\eps)$.
(This requires~$\qq>0$).
Then because for any finite~$\NK$, $\sum_{\km\le\NK} \km^{-2\qq}\lvert\hat\fw_\kv(\time)\rvert^2\rightarrow 0$ as~$\time\rightarrow\infty$, there exists~$\T(\eps) < \infty$ such that~$\sum_{\km\le\NK(\eps)} \km^{-2\qq}\lvert\hat\fw_\kv(\time)\rvert^2 < \eps/2$ for all~$\time>\T(\eps)$.
From~\eqref{eq:laststep} we conclude that~$\norm{\fw}_{\Sobo^{-\qq}}^2 < \eps$ for all~$\time>\T(\eps)$, which proves the result.

\end{document}